\documentclass[fleqn,10pt]{wlscirep}
\usepackage[utf8]{inputenc}
\usepackage[T1]{fontenc}
\usepackage{comment}
\usepackage{booktabs}
\usepackage{bookmark} 

\usepackage{hyperref, bookmark}
\hypersetup{
  pdftitle={Travel time optimization on multi-AGV routing by reverse annealing},
  pdfauthor={Renichiro Haba, Masayuki Ohzeki, Kazuyuki Tanaka},
  pdfkeywords = {reverse annealing, quantum annealing, QUBO, automated guided vehicle, centralized, routing algorithm}
}

\title{Travel time optimization on multi-AGV routing by reverse annealing}

\author[1,3,*]{Renichiro Haba}
\author[1,2,3]{Masayuki Ohzeki}
\author[1]{Kazuyuki Tanaka}

\affil[1]{Graduate School of Information Sciences, Tohoku University, Sendai 980-8579, Japan}
\affil[2]{Institute of Innovative Research, Tokyo Institute of Technology, Kanagawa, 226-8503, Japan}
\affil[3]{Sigma-i Co., Ltd., Tokyo 108-0075, Japan}

\affil[*]{renichiro.haba.r6@dc.tohoku.ac.jp}

\newcommand{\dstar}{d^{*}}


\begin{abstract}
\addcontentsline{toc}{section}{Abstract}
Quantum annealing has been actively researched since D-Wave Systems produced the first commercial machine in 2011. Controlling a large fleet of automated guided vehicles is one of the real-world applications utilizing quantum annealing. In this study, we propose a formulation to control the traveling routes to minimize the travel time. We validate our formulation through simulation in a virtual plant and authenticate the effectiveness for faster distribution compared to a greedy algorithm that does not consider the overall detour distance. Furthermore, we utilize reverse annealing to maximize the advantage of the D-Wave's quantum annealer. Starting from relatively good solutions obtained by a fast greedy algorithm, reverse annealing searches for better solutions around them. Our reverse annealing method improves the performance compared to standard quantum annealing alone and performs up to 10 times faster than the strong classical solver, Gurobi. This study extends a use of optimization with general problem solvers in the application of multi-AGV systems and reveals the potential of reverse annealing as an optimizer.
\end{abstract}
\begin{document} 

\flushbottom
\maketitle

\thispagestyle{empty}

\section*{Introduction}
\addcontentsline{toc}{section}{Introduction}
Automated guided vehicles (AGVs) have been widely employed for transporting materials in factories, warehouses, and container terminals to improve flexibility and efficiency of manufacturing and distribution \cite{ullrichAutomatedGuidedVehicle2015, de_ryck_automated_2020}. They move along markers or wires laid on the floor, which form a network. The design of the network layout can be broadly divided into a general and tandem layout. A tandem layout comprises multiple non-overlapping loops and exactly one AGV is assigned to each loop \cite{bozer_tandem_1991}. By contrast, a general layout does not limit the network topology and all paths can be used by any AGV \cite{gaskins_flow_1987,kaspi_optimal_1990}. A general layout leads to higher flexibility and productivity than a tandem layout because AGVs can travel on alternative routes or shortcuts. However, in such a system, multiple AGVs may share the same segment or intersection simultaneously, which could induce collisions or deadlocks depending on their planned routes. To avoid such a situation, AGVs' traveling routes are planned to be collision- and deadlock-free. In addition, the routes should be the shortest possible to minimize transfer time. Many routing algorithms have been examined for various problem settings \cite{qiu_scheduling_2002,fazlollahtabar_mathematical_2015}. In centralized AGV systems, information related to all AGVs, such as their delivery locations and current positions, is available for optimized routing \cite{jose_task_2016, ohzeki_control_2019}.
Although this centralized optimization is a strong approach, it entails considerable computational time depending on the number of AGVs. To mitigate this issue, we focus on formulating the routing problem as a general mathematical form and exploiting existing fast mathematical solvers or meta-heuristics. As one of the solvers, we are interested in D-Wave Advantage, which is specialized in solving quadratic unconstrained binary optimization (QUBO) problems using quantum annealing.

Quantum annealing is a heuristic algorithm to solve QUBO problems by driving binary variables through quantum fluctuations \cite{kadowaki_quantum_1998}. Several well-known combinatorial problems can be encoded into QUBO problems \cite{lucas_ising_2014}. In the ideal procedure, quantum annealing outputs the optimal solution by decreasing the strength of the fluctuation of binary variables. The quantum adiabatic theorem ensures the theoretical assurance of quantum annealing \cite{suzuki_residual_2005, morita_mathematical_2008, ohzeki_quantum_2011}.
Currently, quantum annealing is realized with an actual machine by D-Wave Systems \cite{johnson_quantum_2011}. The machine does not perform the ideal quantum annealing because of the hardware and environmental effects and the solutions are not always optimal, although it rapidly outputs relatively accurate solutions. As the environmental effect cannot be avoided, the quantum annealer is sometimes regarded as a simulator for the quantum many-body dynamics \cite{Bando2020, Bando2021}. Several practical applications of quantum annealer have been presented across various fields, such as finance \cite{rosenberg_solving_2016, orus_forecasting_2019, venturelli_reverse_2019}, traffic \cite{neukart_traffic_2017, hussain_optimal_2020}, logistics \cite{feld_hybrid_2019, ding_implementation_2021}, manufacturing \cite{venturelli_quantum_2016, ohzeki_control_2019}, and marketing \cite{nishimura_item_2019}, as well as in decoding problems \cite{IdeMaximumLikelihoodChannel2020, Arai2021code}. Its potential for solving the optimization problem with inequality constraints has been enhanced \cite{Yonaga2020}, especially in the case that is hard to formulate directly \cite{Koshikawa2021}. The comparative study of quantum annealer has also been performed for benchmark tests to solve the optimization problems \cite{Oshiyama2022}. The quantum effect on the case with multiple optimal solutions has also been discussed \cite{Yamamoto2020, Maruyama2021}. Further, applications of quantum annealing for machine learning for solving optimization problems have been reported \cite{Amin2018,Kumar2018,Adachi2015,benedetti2016estimation, Arai2021,Sato2021}.

The above-mentioned studies have presented a routing algorithm for multiple AGVs to plan their traveling routes for given pairs of source and destination by utilizing a quantum annealer \cite{ohzeki_control_2019}. The routing aims to improve the working rate of AGVs. This objective is translated into the optimization problem to maximize the total travel distance of AGVs with constraints for removing latent collisions. However, as challenges, AGVs move with unnecessary detours, which reduces delivery efficiency. In addition, depending on the system, AGVs’ repeated back and forth movement or along a cycle sometimes lacks efficiency. Therefore, we aim to develop an algorithm to control AGVs more efficiently. In this study, we formulate a QUBO problem to output deconflicted routes closest to the destinations of AGVs by modifying the objective function discussed in the previous study \cite{ohzeki_control_2019}. For a given task for each AGV, we minimize the total detour distance from the shortest paths. To express the detouring distance, we introduce an estimated distance of AGV, which is the travel distance remaining from the end of the path to the destination. By reducing it for all AGVs, we compel them to move ahead to their destination at each iteration. Consequently, we expect this method to accelerate cargo-carrying capabilities by shortening the detour of AGVs.

To optimize the QUBO problem, we employ D-Wave's quantum annealer for solving the QUBO problem. The latest version, that is, the D-Wave Advantage System, has two features---forward and reverse methods for quantum annealing \cite{chancellor_modernizing_2017, ohkuwa_reverse_2018, noauthor_reverse_2017, Kadowaki2019}.
To reveal the effectiveness of applying D-Wave's machine to our optimization problem, we explore both forward and reverse annealing features, which have not been hitherto discussed in AGVs' routing applications. In short, forward annealing performs a global search, starting in a superposition of all possible states with equal weights. By contrast, reverse annealing searches locally around a given classical state. Reverse annealing refines a solution around the given classical state unlike forward annealing, which searches for all possible states equally. For some types of problems, the reverse annealing approach improves the performance by utilizing initial solutions obtained by forward annealing or classical algorithms \cite{venturelli_reverse_2019, golden_reverse_2021}. In this study, we present a reverse annealing method combined with a fast greedy search for solving the QUBO problem. The problem is also solved via a linear programming solver called Gurobi \cite{noauthor_gurobi_2020}. To evaluate the performance, we measure the time-to-solution benchmark for each solver.

The remainder of this paper is organized as follows: In the next section, we first review the previous method for the control of AGVs with less congestion. Next, we formulate the control of AGVs with fewer detours as the QUBO problem by modifying the objective function. Then, we present the optimization method that utilizes reverse annealing.
To obtain the initial solution, we construct the greedy algorithm for the preprocessing of the reverse annealing method. In the following section, we evaluate our method through a simulation in a virtual plant. To validate the effectiveness of D-Wave's machine, we compare the results of the obtained solutions by each solver. In the final section, we summarize our study and discuss how future studies can improve the performance of solutions to this problem.

\section*{Method}
\addcontentsline{toc}{section}{Method}
In this section, we introduce our method for controlling AGVs using a quantum annealer. First, we review the study by Ohzeki \textit{et al}. \cite{ohzeki_control_2019}. We briefly explain the routing algorithm and definition of the optimization problem and specify its issues. Then, we redefine the optimization problem to minimize the total journey time of AGVs. The optimization problem is relaxed to a QUBO form, which is heuristically solved by a quantum annealer. Finally, we describe how to run reverse annealing for the QUBO problem.

Assume that loads assigned to AGVs and their start and target positions are given. In the routing algorithm, we consider finding a route for each AGV to carry its load to its destination node from the source node in the network.
We call a pair of source and destination nodes a task. A static routing algorithm determines a route for each task in advance, whereas a dynamic algorithm decides the route based on real-time information. A dynamic algorithm is advantageous as the routes are flexibly adjusted to reduce congestion and travel time.

For using a quantum annealer for AGVs' routing, a dynamic algorithm with QUBO optimization has been proposed\cite{ohzeki_control_2019}. The algorithm focuses on improving the working rate of AGVs. For dynamically updated tasks, the algorithm provides a good route for each AGV during a fixed period $T$. First, for each AGV, we generate multiple dissimilar candidate routes from its current node $v_0$ to destination $v_t$ and shorten them to enable reaching within $T$ s. Each candidate route is expressed as a list of nodes such that it connects $x$ to $t$ on the network.
Second, we divide each candidate route into lists of segments to implement AGVs' stopping on the route. For example, if one candidate route consists of a path $(v_0, v_1, v_t)$, then we split it into lists $(v_0, v_1, v_t)$, $(v_0, v_1)$, and $(v_0)$, which represent moving the AGV two segments ahead, one segment ahead, and stopping, respectively. Note that in this algorithm, AGVs cannot go ahead during $T$ second once they stop. 
Third, for all AGVs, we find the routes that maximize the working rate from the candidate routes. Finally, we move AGVs along the selected routes in $T$ second and iterate the above procedure.

The procedure of choosing the best route is regarded as an optimization problem to maximize AGVs' working rate. The optimization problem comprises an objective function to be maximized or minimized with constraints. To represent the working rate, the objective function is defined as the total distance that AGVs move in the network. 
By maximizing the objective function, the total distance of routes is maximized and the solution leads to the highest working rate in the AGVs' system.

However, we have two issues with the objective function of the algorithm.
One is inducing a longer detour distance, while the desired AGV's route should be the shortest possible to ensure fast delivery. The other is the possible suppressing of the system because of AGV's continuous back and forth movement in certain areas. To overcome these challenges, we present a new definition of the optimization problem to minimize travel time. 

The travel time is proportional to the route distance and can be directly used for the objective function. However, recalling that we use the divided routes and not all of them reach the destination nodes, the length of the shortest route equals zero and that causes all AGVs to stop. In this study, we introduce the \textit{remaining distance} instead of the route distance. For a given $i$-th AGV and $j$-th route, the remaining distance $\dstar_{i,j}$ is the length of the shortest path from the end node of the route to its destination. By choosing the shortest route in terms of the remaining distance, AGVs move to nodes closest to their destination nodes for each iteration in the algorithm. Therefore, the algorithm greedily minimizes AGVs’ travel time to complete their tasks. Let $q_{ij}$ be a binary variable that takes a value $1$ if the $i$-th AGV moves on the $j$-route and $0$ otherwise. Then, the objective function is defined as
\begin{equation}
    \label{Prop_Obj}
    f_{\rm{obj}}(\mathbf{q}) = \sum_{i=1}^{N} \sum_{j=1}^{M_i} \dstar_{i, j} q_{i,j}
\end{equation}
where $N$ is the number of AGVs and $M_i$ is the number of candidate routes for the $i$-th AGV.

To safely control AGVs, collisions on roads or at intersections must be avoided.
Deconfliction can be implemented by slowing down or stopping the AGVs after the traveling routes are arranged; however, we perform it simultaneously during the optimization.
We utilize a similar technique discussed in the previous study to define constraints for collision avoidance \cite{ohzeki_control_2019}.
We allow at most one AGV to occupy each line segment and intersection.
We define $F_{i,j,t,e}$ such that $F_{i,j,t,e} = 1$ if the segment $e$ is occupied with the $j$-th route of the $i$-th AGV at time $t$ and $F_{i,j,t,e} = 0$ otherwise. Note that if an AGV is on a segment $uv$, we consider any segment coming to $v$ as occupied.
For any edge $e$ and time $t$, $\sum_{i=1}^{N} \sum_{j=1}^{M_i} F_{i,j,t,e} q_{i,j}$ has to be at most one.

Now we define a mixed integer linear programming (MILP) problem as
\begin{equation}
\begin{split}
    \min _{\mathbf{q}} & ~~~~ \sum_{i=1}^{N} \sum_{j=1}^{M_i} \dstar_{i, j} q_{i,j} \\
    \rm{subject~to} & ~~~~ \sum_{j=1}^{M_i} q_{i,j} = 1,  ~~\forall i \in \{1, 2, \ldots, N \} \\
    & ~~~~ \sum_{i=1}^{N} \sum_{j=1}^{M_i} F_{i,j,t,e} q_{i,j} \le 1,  ~~\forall e \in E, ~~\forall t \in \{1, 2, \ldots , T\}
    \label{MILP}
\end{split}
\end{equation}
where $E$ represents a set of edges in the network.
The first constraint ensures that each AGV picks a single route and the second one ensures deconflicted routes. In general MILP problems, finding the exact solution quickly is difficult, although classical methods such as branch-and-bound algorithm are utilized to traverse the solution space efficiently \cite{land_automatic_1960}.

Next, we transform the MILP problem into a QUBO problem to utilize quantum annealing. By using the penalty method for the equality and inequality constraints in problem \eqref{MILP}, the cost function to be minimized in the QUBO problem can be expressed as:
\begin{equation}
    f_{\rm{QUBO}} (\mathbf{q}) = \sum_{i=1}^{N} \sum_{j=1}^{M_i} \dstar_{i, j} q_{i,j} + a \sum_{i=1}^{N} \left( \sum_{j=1}^{M_i} q_{i,j} - 1\right)^2  + b \sum_{e \in E} \sum_{t=1}^{T}\left(\sum_{i=1}^{N} \sum_{j=1}^{M_i} F_{i,j,t,e} q_{i,j} - \frac{1}{2} \right)^2
    \label{qubo}
\end{equation}
where $a>0$ and $b>0$ denote penalty parameters that are tuned to ensure that the constraints are satisfied with the minimum energy. The second and third terms represent the equality and inequality constraints in problem \eqref{MILP}, respectively.
Thus, we have the following quadratic form with a matrix $Q$ by transforming the function \eqref{qubo}.
\begin{equation}
    f_{\rm{QUBO}} (\mathbf{q}) = \mathbf{q}^\mathrm{T} Q \mathbf{q} + \mathrm{Const.}
    \label{Q}
\end{equation}
General QUBO and MILP problems are known to be hard to strictly optimize.
As QUBO problems do not have any constraints, the size of their feasible solution space tends to be fairly larger than that of MILP problems for the same problem size. Thus, heuristic algorithms are widely applied for quickly obtaining an adequate solution.
In this study, we employ D-Wave's quantum annealer to solve the QUBO problem heuristically.

QUBO problems are equivalently converted to minimizing the energy of the Ising model, for which the Hamiltonian is expressed as:
\begin{equation}
    H_0 (\mathbf{\sigma}) = - \sum_{i} h_i \sigma_i - \sum_{i<j} J_{ij} \sigma_i \sigma_j \ 
    \label{Ising}
\end{equation}
where $\sigma_i \in \{-1, 1\}$ denotes a spin variable.

In quantum annealing, the quantum fluctuation is utilized to effectively find solutions of Ising models. The D-Wave machine operates the quantum system with superconducting qubits in a transverse field and its Hamiltonian is expressed as:
\begin{equation}
    \hat{H} (s) = - A(s) \sum_i \hat{\sigma}_i^x + B(s) \hat{H}_0\ 
    \label{eq:QA_hamiltonian}
\end{equation}
where $\hat{\sigma}_i^x$ is the $x$-component of Pauli matrices and $\hat{H}_0$ is a Hamiltonian attained by replacing each spin variable $\sigma_i$ with the $z$-component of Pauli matrices $\hat{\sigma}_i^z$.
The Hamiltonian \eqref{eq:QA_hamiltonian} is controlled by a predetermined annealing schedule with a time-dependent parameter $0 \le s \le 1$.
The functions $A(s)$ and $B(s)$ control the magnitude of transverse field to satisfy $A(0) \gg B(0)$ and $A(1) \ll B(1)$.
At $A(0) \gg B(0)$, the qubits have the trivial ground state with a uniform superposition of all possible states. At $A(1) \ll B(1)$, the system has a classical state equivalent to the spin variables.

Depending on the annealing schedule, quantum annealing is classified into forward and reverse annealing. Forward annealing starts with $s = 0$ and gradually increases to $s = 1$ in the predetermined annealing time. By gradually decreasing the magnitude of the transverse field, the qubits are dephased into a nontrivial classical state of the Ising system. In an ideal procedure of forward annealing, the optimal state of the Ising model is known to be attained. However, D-Wave's quantum annealer couples to an open system that is affected by its environment, which leads to a limited optimization performance.

Unlike forward annealing, which is initialized with the full superposition, reverse annealing starts with a classical state and searches for a better solution in its vicinity. In reverse annealing, the system starts with $s=1$ and gradually decreases $s$ to $s=1-r$, where $0 < r < 1$ is the reversal distance, which determines the strength of transverse field. Pausing the schedule at $s=1-r$ for a certain period exploits the thermal relaxation in a low-temperature bath and enhances the optimization performance. After the pausing, the system is annealed forward to $s=1$ and the qubits are dephased into a classical state.

In this study, we employ reverse annealing mainly for two reasons. One is to improve the performance of optimizations by combining a fast classical heuristic algorithm. The other is to ensure the feasibility of solutions. Both forward and reverse annealing have no assurance of the output precision and infeasible solutions could be obtained, which induce fatal errors in manufacturing systems. Thus, an alternative technique to find feasible solutions is necessary for practical use, which can be utilized if the annealing method fails to find feasible solutions. Furthermore, reverse annealing provides an opportunity to refine such feasible solutions and exceed the performance of forward annealing alone.

We introduce a greedy algorithm to derive feasible solutions and utilize them as initial states for reverse annealing. The simplified flow of the algorithm is as follows.
\begin{enumerate}
  \item Allocate the shortest route to each vehicle.
  \item For any one of the two vehicles that collide, assign the next route.
  \item Iterate 2 until all collisions are removed.
\end{enumerate}
In the algorithm, the selections of any two conflicting routes rely on the calculation of an impact value, which is the number of conflicting AGVs for their next candidate routes. One route with a smaller impact value is dismissed and the other route with a larger impact value is selected. By exploring the candidate routes in the ascending orders of the remaining distance, the algorithm greedily searches the solutions. As the candidate routes are separated to stop on the way, the algorithm always finds conflict-free routes. This algorithm runs much faster than solving the MILP or QUBO problems. Specifically, the number of iterations in the algorithm is $O(nm)$, where $n$ and $m$ are the number of AGVs and candidate routes, respectively.

The greedy algorithm does not consider any detour distance in the selection of routes and aims to quickly find the deconflicted routes, which results in exploiting the high working rate. This feature enables us to imitate conditions similar to those in the previous study. The working rate of AGVs is also preferred to be high in the AGV systems to avoid traffic congestion. In the next section, we compare the algorithms in terms of the working rate and carrying efficiency.

\begin{figure}[tbp]
  \centering
  \includegraphics[width=12cm]{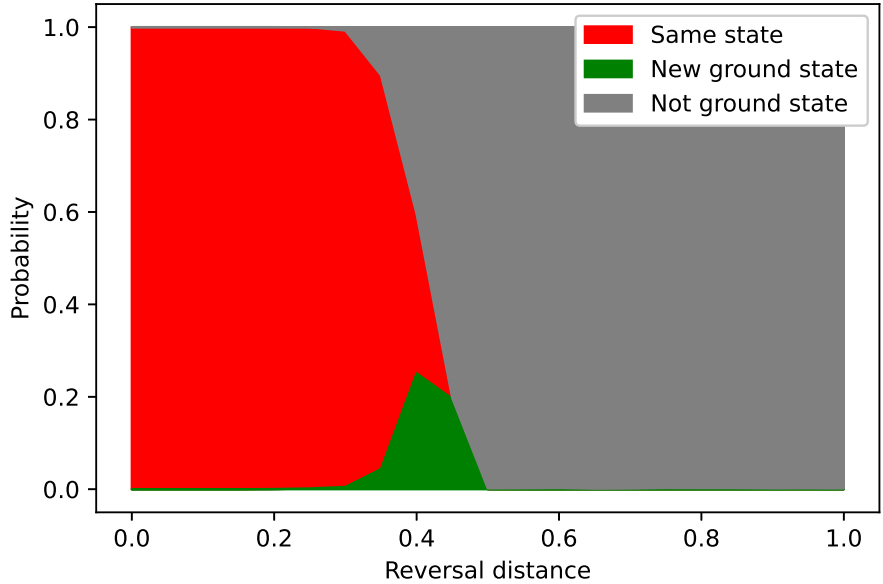}
  \caption{\textbf{Calibration of reversal distance.} We performed reverse annealing on 10 random nontrivial QUBO problems that appeared while running the algorithm with $20$ AGVs and analyzed the energies of each of the $1000$ samples. For a given reversal distance, the height of the red, green, and gray areas represents the mean probability of attaining the same state, the ground state, and the other state, respectively.  }
  \label{fig:reversal_distance}
\end{figure}

\begin{figure}[tbp]
  \centering
  \includegraphics[width=8cm]{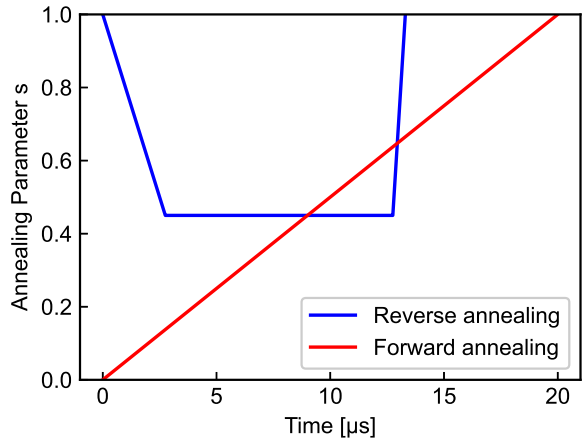}
  \caption{\textbf{Annealing schedule of forward and reverse annealing.} The red and blue lines indicate the chronological changes in annealing parameter $s$ in forward and reverse annealing, respectively. }
  \label{fig:annealing_schedule}
\end{figure}

To exploit the performance of reverse annealing, we maintain the annealing schedule properly. The most dominant parameter is reversal distance $r$.
If it is too small, the solution stays in the initial state and if it is too large, the probability of achieving a global minimum by performing the global search equivalently as forward annealing is low. To find the reversal distance with a high possibility to output optimal solutions, for 10 randomly chosen QUBO matrices during the algorithm for controlling $20$ AGVs, we analyzed the solutions for different reversal distances. The result is illustrated in Fig. \ref{fig:reversal_distance}. Obtaining the same initial state had a high probability with a small reversal distance, which gradually decreases as reversal distance increases. When reversal distance is too large, reverse annealing fails to find optimal states in most cases. Overall, the count of obtaining optimal solutions peaks around $r = 0.45$. Thus, we find the reversal distance $0.45$ as the probability of obtaining the ground states peaked around the value and the probability of obtaining the same states is small. Finally, we have a calibrated reverse annealing schedule shown in Fig. \ref{fig:annealing_schedule}. The default annealing time of D-Wave's quantum annealer is $20$ µs and we use this setting in forward annealing.
We set the annealing time in reverse annealing to $13.3$ µs with a pause of $10$ µs, which is shorter than that in forward annealing.

\section*{Results}
\begin{figure}[tbp]
  \centering
  \includegraphics[width=15cm]{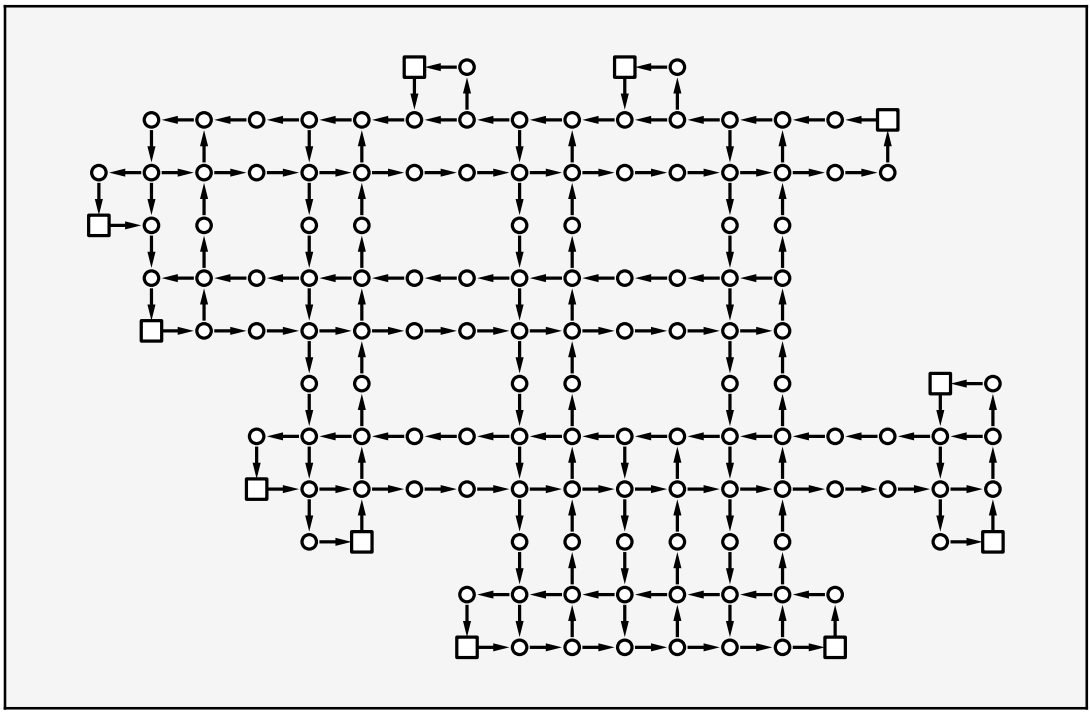}
  \caption{\textbf{Virtual plant we used in this study.} The network shows the roads on which AGVs can move. Each circle represents a node, which is an intersection or a mid-point. Each arrow represents the directed connection between two nodes. The length of each arrow is one meter. The square nodes indicate the pick-up or drop-off points that can be the source and destination of AGVs. In the plant, 20 AGVs are active on their task. }
  \label{fig:graph}
\end{figure}

\addcontentsline{toc}{section}{Results}
In this section, we report the results of the dynamic algorithm solving the QUBO problem at each time. To verify the effectiveness of the algorithm, we simulate the AGV system in a simple virtual plant, as illustrated in Fig. \ref{fig:graph}. In the plant, $20$ AGVs are active and they are provided a list of tasks that have to be completed at the earliest. In fixed simulation time, the number of completed tasks corresponds to the efficiency of carrying, which is also evaluated by the total travel time to finish the fixed number of tasks. The speed of each AGV is set to $0.5$ m/s. 

We iterate the algorithm $500$ times and observe the number of tasks completed and the working rate. In the simulation, different solvers are utilized to obtain the solution to each QUBO problem, specifically, greedy algorithm, Gurobi, forward annealing, and reverse annealing. As mentioned above, the greedy algorithm attempts to avoid congestion by reducing the number of AGVs stopping. Thus, we regard it as similar to the previous algorithm and employ it to compare with our method. For forward and reverse annealing, we set the number of samples to $1,000$ and the state with the lowest energy is chosen in the routing algorithm. The greedy algorithm and Gurobi are operated on Intel(R) Core(TM) i7-8569U CPU.

The results of the simulation are presented in Table \ref{table:simulation}.
At each optimization, Gurobi always outputs the optimal solution and leads to the highest performance in completed tasks. Although the working rate of the greedy algorithm was higher than that of Gurobi, the completed tasks of the greedy algorithm were less than that of Gurobi. This is because AGVs move with unnecessary detours in the greedy algorithm, which indicates that travel time is reduced by solving our optimization problem. In our case, forward annealing performed worst both in the completed tasks and working rate because of failing to find low-energy solutions. By contrast, reverse annealing outperformed forward annealing and resulted in almost the same scores as those of Gurobi. To conclude, the optimization of the QUBO problem \ref{qubo} is effective in realizing a time-efficient multi-AGV system as well as solving the MILP problem. In addition, Gurobi and reverse annealing are comparable as solvers to control $20$ AGVs.

\begin{table}[tbp]
    \caption{\textbf{Comparison of algorithms and solvers.} The mean number of completed tasks for 500 iterations and the working rate of AGVs obtained by the greedy algorithm, Gurobi, forward annealing, and reverse annealing. Data are drawn from 10 identical experiments. The standard deviation is shown on the right hand side of the plus-minus sign. }
    \label{table:simulation}
    \centering
    \begin{tabular}{cllll} 
        \toprule
        & Greedy & Gurobi & Forward annealing & Reverse annealing \\ \midrule
        Completed tasks & $603.0$ & $629.0$ & $542.1 \pm{2.3}$ & $627.4 \pm{1.9}$\\
        Working rate (\%) & $98.9$ & $97.8$ & $84.9 \pm{0.2}$ & $97.8 \pm{0.1}$\\ \bottomrule
    \end{tabular}
\end{table}

\begin{figure}[p]
  \centering
  \includegraphics[width=14cm]{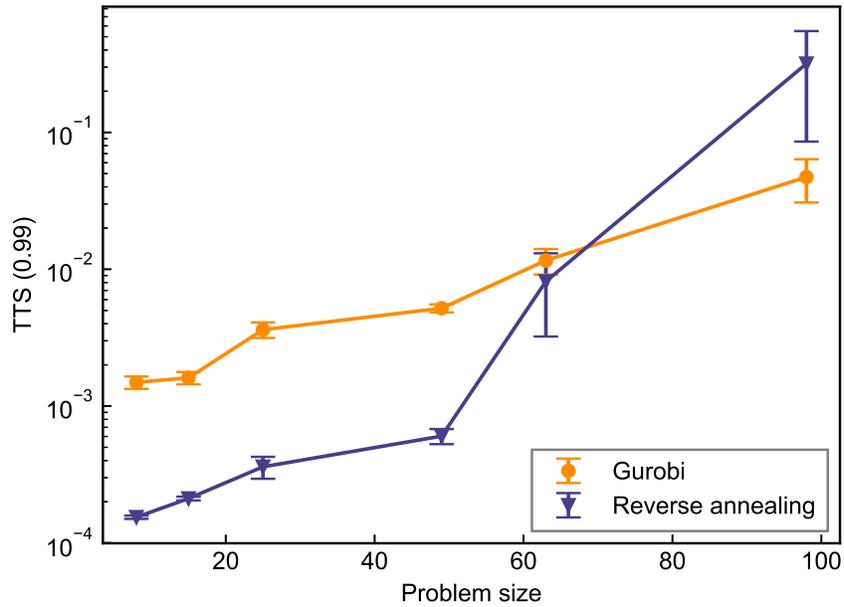}
  \caption{\textbf{Comparison of TTS performance} The orange circle and purple triangle show mean value of TTS(0.99) of Gurobi and reverse annealing, respectively. The error bars indicate the standard error.}
  \label{fig:TTS}
\end{figure}

To compare the performance of solvers, we benchmark time-to-solutions (TTS), defined as:
\begin{equation}
    {\rm TTS} (p) = t_c \frac{\log (1-p)}{\log (1-p_s)},
\end{equation}
where $p$ is the probability of obtaining the optimal solution at least once with a fixed number of trials, $p_s$ is the probability of obtaining the optimal solution with a single trial, and $t_c$ is the computational time for a single trial. For example, TTS(0.99) shows the estimated amount of time to obtain the optimal solution with a $99$ \% chance. For the computation time of reverse annealing, we use annealing time during access time on the quantum processing unit. The annealing time is the total amount of time to complete a given annealing schedule. For reverse annealing, we apply the annealing schedule as shown in Fig. \ref{fig:annealing_schedule} and the annealing time of a single anneal is $13.3$ µs. To explore the problem size scalability of the solvers, we measure TTS for different numbers of AGVs.
The problem size is the summation of the number of candidate routes for each AGV. Note that the problem size differs in the algorithm because of separating routes to let AGVs wait. We calculate the TTS(0.99) for $10$ QUBO problems appearing in the simulation. The number of samples in reverse annealing is set to $10,000$. The result is depicted in Fig. \ref{fig:TTS}. We plot the wall-clock time for Gurobi to obtain the optimal solution instead of TTS. Reverse annealing outperforms Gurobi with almost $10$ times shorter time when the problem size is small. As the problem size increased, reverse annealing needed a longer time to obtain the optimal solution, and Gurobi outperformed reverse annealing. Reverse annealing failed to find the optimal solution when the problem size was over $100$ and benchmarking TTS was impossible. We assume this is because initial solutions obtained by the greedy algorithm become worse for larger problems.

\begin{figure}[p]
  \centering
  \includegraphics[width=12cm]{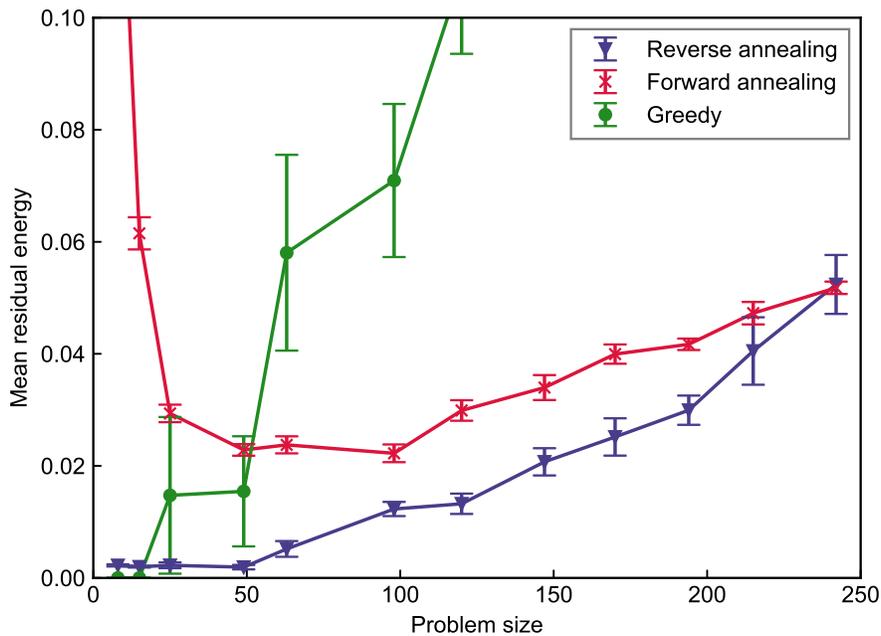}
  \caption{\textbf{Comparison of residual energy.} The purple triangle, red cross, and green circle show mean residual energy of samples on 10 problems solved by reverse annealing, forward annealing, and the greedy algorithm, respectively. The error bars indicate the standard error. }
  \label{fig:Residual Energy}
\end{figure}

To investigate the effect of initial solutions, we evaluate residual energy defined as:
\begin{equation}
    E_{\rm res} = \frac {\langle E \rangle - E_{\rm min}}{E_{\rm min}}
\end{equation}
where $\langle E \rangle$ is the mean energy of samples and $E_{\rm min}$ is the energy of the optimal solution. The residual energy corresponds to the closeness between the mean and minimum energy. 
We obtain 10,000 samples by forward and reverse annealing. The result is depicted in Fig. \ref{fig:Residual Energy}. The greedy algorithm always leads to a unique solution, and we plot the fraction of its energy over the minimum one. The greedy algorithm has extremely high residual energy as problem size increases. When the problem size is less than $50$, the residual energy of forward annealing is higher than that of the greedy algorithm, which indicates that the greedy algorithm performs better than forward annealing. By contrast, reverse annealing resulted in lower residual energy even for small problems. In reverse annealing, the residual energy tends to be higher but still less than forward annealing as the problem size increases up to $215$. For $250$ variables, reverse annealing has higher residual energy than forward annealing and may turn worse for problem sizes larger than $250$. 

\section*{Discussion}
\addcontentsline{toc}{section}{Discussion}
In this study, we formulated a QUBO problem to control multiple AGVs' routes with minimized detours for a given period. Using a simulation, we demonstrated that our algorithm yields time-efficient routing. However, controlling AGVs as simulated is not always possible because of unpredictable events such as human errors and communication delays.
To test our methods under such practical conditions and realize robust routing algorithms is interesting.

As an optimization method, we utilized reverse annealing with initial states obtained by a fast greedy algorithm. We confirmed that our reverse annealing method has the potential to exceed forward annealing alone or even a classical commercial solver, Gurobi. Practically, reverse annealing performed up to 10 times faster than Gurobi to obtain optimal solutions for small problem sizes, whereas forward annealing suffered in finding optimal solutions. 
We reckon this unexpected behavior of D-Wave Advantage 1.1 is due to its unique characteristic. According to the technical report by D-Wave Systems, such behavior is not announced and the system seems to perform as expected \cite{mcgeochDWaveAdvantageSystem}. Thus, D-Wave Advantage 1.1 may have a weakness for the types of problems that we seek to solve.

By contrast, reverse annealing incorporates the benefits of both the greedy algorithm and quantum annealing and seems to have stable performance. One possible approach to improve our reverse annealing methods is switching the initial algorithm from the greedy algorithm to forward annealing depending on their performance. The residual energy of the greedy algorithm is considerably higher than the minimum one, although reverse annealing corrects the initial state to a certain extent. Choosing the right initial solver for large problems will improve the performance of reverse annealing. One choice of the approximate solvers can be a mean-field approximation and its variants \cite{Ohzeki2019JPSJ}. The performance of reverse annealing with the mean-field approximation has been investigated \cite{Arai2021code}. Employing approximation solvers with a theoretical guarantee for setting initial conditions on reverse annealing can enable us to investigate the performance and design a theoretical method for assessing its limitation. As demonstrated in our study, although it is just one of the optimization problems, the performance of reverse annealing with the greedy algorithm is comparable with that of the commercial best optimizer. By assessing various combinations of reverse annealing and approximate solvers, we may find a new classical-quantum hybrid scheme to solve large-scale optimization problems.

\bibliography{Reference}

\begin{thebibliography}{10}
\urlstyle{rm}
\expandafter\ifx\csname url\endcsname\relax
  \def\url#1{\texttt{#1}}\fi
\expandafter\ifx\csname urlprefix\endcsname\relax\def\urlprefix{URL }\fi
\expandafter\ifx\csname doiprefix\endcsname\relax\def\doiprefix{DOI: }\fi
\providecommand{\bibinfo}[2]{#2}
\providecommand{\eprint}[2][]{\url{#2}}

\bibitem{ullrichAutomatedGuidedVehicle2015}
\bibinfo{author}{Ullrich, G.}
\newblock \emph{\bibinfo{title}{Automated {{Guided Vehicle Systems}}}}
  (\bibinfo{publisher}{Springer}, \bibinfo{year}{2015}).

\bibitem{de_ryck_automated_2020}
\bibinfo{author}{De~Ryck, M.}, \bibinfo{author}{Versteyhe, M.} \&
  \bibinfo{author}{Debrouwere, F.}
\newblock \bibinfo{journal}{\bibinfo{title}{Automated guided vehicle systems,
  state-of-the-art control algorithms and techniques}}.
\newblock {\emph{\JournalTitle{Journal of Manufacturing Systems}}}
  \textbf{\bibinfo{volume}{54}}, \bibinfo{pages}{152--173},
  \doiprefix\url{http://doi.org/10.1016/j.jmsy.2019.12.002}
  (\bibinfo{year}{2020}).

\bibitem{bozer_tandem_1991}
\bibinfo{author}{BOZER, Y.~A.} \& \bibinfo{author}{SRINIVASAN, M.~M.}
\newblock \bibinfo{journal}{\bibinfo{title}{Tandem {Configurations} for
  {Automated} {Guided} {Vehicle} {Systems} and the {Analysis} of {Single}
  {Vehicle} {Loops}}}.
\newblock {\emph{\JournalTitle{IIE Transactions}}}
  \textbf{\bibinfo{volume}{23}}, \bibinfo{pages}{72--82},
  \doiprefix\url{http://doi.org/10.1080/07408179108963842}
  (\bibinfo{year}{1991}).

\bibitem{gaskins_flow_1987}
\bibinfo{author}{GASKINS, R.~J.} \& \bibinfo{author}{TANCHOCO, J. M.~A.}
\newblock \bibinfo{journal}{\bibinfo{title}{Flow path design for automated
  guided vehicle systems}}.
\newblock {\emph{\JournalTitle{International Journal of Production Research}}}
  \textbf{\bibinfo{volume}{25}}, \bibinfo{pages}{667--676},
  \doiprefix\url{http://doi.org/10.1080/00207548708919869}
  (\bibinfo{year}{1987}).

\bibitem{kaspi_optimal_1990}
\bibinfo{author}{KASPI, M.} \& \bibinfo{author}{TANCHOCO, J. M.~A.}
\newblock \bibinfo{journal}{\bibinfo{title}{Optimal flow path design of
  unidirectional {AGV} systems}}.
\newblock {\emph{\JournalTitle{International Journal of Production Research}}}
  \textbf{\bibinfo{volume}{28}}, \bibinfo{pages}{1023--1030},
  \doiprefix\url{http://doi.org/10.1080/00207549008942772}
  (\bibinfo{year}{1990}).

\bibitem{qiu_scheduling_2002}
\bibinfo{author}{Qiu, L.}, \bibinfo{author}{Hsu, W.-J.},
  \bibinfo{author}{Huang, S.-Y.} \& \bibinfo{author}{Wang, H.}
\newblock \bibinfo{journal}{\bibinfo{title}{Scheduling and routing algorithms
  for {AGVs}: {A} survey}}.
\newblock {\emph{\JournalTitle{International Journal of Production Research}}}
  \textbf{\bibinfo{volume}{40}}, \bibinfo{pages}{745--760},
  \doiprefix\url{http://doi.org/10.1080/00207540110091712}
  (\bibinfo{year}{2002}).

\bibitem{fazlollahtabar_mathematical_2015}
\bibinfo{author}{Fazlollahtabar, H.}, \bibinfo{author}{Saidi-Mehrabad, M.} \&
  \bibinfo{author}{Balakrishnan, J.}
\newblock \bibinfo{journal}{\bibinfo{title}{Mathematical optimization for
  earliness/tardiness minimization in a multiple automated guided vehicle
  manufacturing system via integrated heuristic algorithms}}.
\newblock {\emph{\JournalTitle{Robotics and Autonomous Systems}}}
  \textbf{\bibinfo{volume}{72}}, \bibinfo{pages}{131--138},
  \doiprefix\url{http://doi.org/10.1016/j.robot.2015.05.002}
  (\bibinfo{year}{2015}).

\bibitem{jose_task_2016}
\bibinfo{author}{Jose, K.} \& \bibinfo{author}{Pratihar, D.~K.}
\newblock \bibinfo{journal}{\bibinfo{title}{Task allocation and collision-free
  path planning of centralized multi-robots system for industrial plant
  inspection using heuristic methods}}.
\newblock {\emph{\JournalTitle{Robotics and Autonomous Systems}}}
  \textbf{\bibinfo{volume}{80}}, \bibinfo{pages}{34--42},
  \doiprefix\url{http://doi.org/10.1016/j.robot.2016.02.003}
  (\bibinfo{year}{2016}).

\bibitem{ohzeki_control_2019}
\bibinfo{author}{Ohzeki, M.}, \bibinfo{author}{Miki, A.},
  \bibinfo{author}{Miyama, M.~J.} \& \bibinfo{author}{Terabe, M.}
\newblock \bibinfo{journal}{\bibinfo{title}{Control of {Automated} {Guided}
  {Vehicles} {Without} {Collision} by {Quantum} {Annealer} and {Digital}
  {Devices}}}.
\newblock {\emph{\JournalTitle{Frontiers in Computer Science}}}
  \textbf{\bibinfo{volume}{1}}, \bibinfo{pages}{9},
  \doiprefix\url{http://doi.org/10.3389/fcomp.2019.00009}
  (\bibinfo{year}{2019}).

\bibitem{kadowaki_quantum_1998}
\bibinfo{author}{Kadowaki, T.} \& \bibinfo{author}{Nishimori, H.}
\newblock \bibinfo{journal}{\bibinfo{title}{Quantum annealing in the transverse
  {Ising} model}}.
\newblock {\emph{\JournalTitle{Physical Review E}}}
  \textbf{\bibinfo{volume}{58}}, \bibinfo{pages}{5355--5363},
  \doiprefix\url{http://doi.org/10.1103/PhysRevE.58.5355}
  (\bibinfo{year}{1998}).

\bibitem{lucas_ising_2014}
\bibinfo{author}{Lucas, A.}
\newblock \bibinfo{journal}{\bibinfo{title}{Ising formulations of many {NP}
  problems}}.
\newblock {\emph{\JournalTitle{Frontiers in Physics}}}
  \textbf{\bibinfo{volume}{2}}, \bibinfo{pages}{5},
  \doiprefix\url{http://doi.org/10.3389/fphy.2014.00005}
  (\bibinfo{year}{2014}).

\bibitem{suzuki_residual_2005}
\bibinfo{author}{Suzuki, S.} \& \bibinfo{author}{Okada, M.}
\newblock \bibinfo{journal}{\bibinfo{title}{Residual {Energies} after {Slow}
  {Quantum} {Annealing}}}.
\newblock {\emph{\JournalTitle{Journal of the Physical Society of Japan}}}
  \textbf{\bibinfo{volume}{74}}, \bibinfo{pages}{1649--1652},
  \doiprefix\url{http://doi.org/10.1143/JPSJ.74.1649} (\bibinfo{year}{2005}).

\bibitem{morita_mathematical_2008}
\bibinfo{author}{Morita, S.} \& \bibinfo{author}{Nishimori, H.}
\newblock \bibinfo{journal}{\bibinfo{title}{Mathematical foundation of quantum
  annealing}}.
\newblock {\emph{\JournalTitle{Journal of Mathematical Physics}}}
  \textbf{\bibinfo{volume}{49}}, \bibinfo{pages}{125210},
  \doiprefix\url{http://doi.org/10.1063/1.2995837} (\bibinfo{year}{2008}).

\bibitem{ohzeki_quantum_2011}
\bibinfo{author}{Ohzeki, M.} \& \bibinfo{author}{Nishimori, H.}
\newblock \bibinfo{journal}{\bibinfo{title}{Quantum {Annealing}: {An}
  {Introduction} and {New} {Developments}}}.
\newblock {\emph{\JournalTitle{Journal of Computational and Theoretical
  Nanoscience}}} \textbf{\bibinfo{volume}{8}}, \bibinfo{pages}{963--971},
  \doiprefix\url{http://doi.org/10.1166/jctn.2011.1776963}
  (\bibinfo{year}{2011}).

\bibitem{johnson_quantum_2011}
\bibinfo{author}{Johnson, M.~W.} \emph{et~al.}
\newblock \bibinfo{journal}{\bibinfo{title}{Quantum annealing with manufactured
  spins}}.
\newblock {\emph{\JournalTitle{Nature}}} \textbf{\bibinfo{volume}{473}},
  \bibinfo{pages}{194--198}, \doiprefix\url{http://doi.org/10.1038/nature10012}
  (\bibinfo{year}{2011}).

\bibitem{Bando2020}
\bibinfo{author}{Bando, Y.} \emph{et~al.}
\newblock \bibinfo{journal}{\bibinfo{title}{Probing the universality of
  topological defect formation in a quantum annealer: Kibble-zurek mechanism
  and beyond}}.
\newblock {\emph{\JournalTitle{Phys. Rev. Research}}}
  \textbf{\bibinfo{volume}{2}}, \bibinfo{pages}{033369},
  \doiprefix\url{http://doi.org/10.1103/PhysRevResearch.2.033369}
  (\bibinfo{year}{2020}).

\bibitem{Bando2021}
\bibinfo{author}{Bando, Y.} \& \bibinfo{author}{Nishimori, H.}
\newblock \bibinfo{journal}{\bibinfo{title}{Simulated quantum annealing as a
  simulator of nonequilibrium quantum dynamics}}.
\newblock {\emph{\JournalTitle{Phys. Rev. A}}} \textbf{\bibinfo{volume}{104}},
  \bibinfo{pages}{022607},
  \doiprefix\url{http://doi.org/10.1103/PhysRevA.104.022607}
  (\bibinfo{year}{2021}).

\bibitem{rosenberg_solving_2016}
\bibinfo{author}{Rosenberg, G.} \emph{et~al.}
\newblock \bibinfo{journal}{\bibinfo{title}{Solving the {Optimal} {Trading}
  {Trajectory} {Problem} {Using} a {Quantum} {Annealer}}}.
\newblock {\emph{\JournalTitle{IEEE Journal of Selected Topics in Signal
  Processing}}} \textbf{\bibinfo{volume}{10}}, \bibinfo{pages}{1053--1060},
  \doiprefix\url{http://doi.org/10.1109/JSTSP.2016.2574703}
  (\bibinfo{year}{2016}).

\bibitem{orus_forecasting_2019}
\bibinfo{author}{Orús, R.}, \bibinfo{author}{Mugel, S.} \&
  \bibinfo{author}{Lizaso, E.}
\newblock \bibinfo{journal}{\bibinfo{title}{Forecasting financial crashes with
  quantum computing}}.
\newblock {\emph{\JournalTitle{Physical Review A}}}
  \textbf{\bibinfo{volume}{99}}, \bibinfo{pages}{060301},
  \doiprefix\url{http://doi.org/10.1103/PhysRevA.99.060301}
  (\bibinfo{year}{2019}).

\bibitem{venturelli_reverse_2019}
\bibinfo{author}{Venturelli, D.} \& \bibinfo{author}{Kondratyev, A.}
\newblock \bibinfo{journal}{\bibinfo{title}{Reverse quantum annealing approach
  to portfolio optimization problems}}.
\newblock {\emph{\JournalTitle{Quantum Machine Intelligence}}}
  \textbf{\bibinfo{volume}{1}}, \bibinfo{pages}{17--30},
  \doiprefix\url{http://doi.org/10.1007/s42484-019-00001-w}
  (\bibinfo{year}{2019}).

\bibitem{neukart_traffic_2017}
\bibinfo{author}{Neukart, F.} \emph{et~al.}
\newblock \bibinfo{journal}{\bibinfo{title}{Traffic {Flow} {Optimization}
  {Using} a {Quantum} {Annealer}}}.
\newblock {\emph{\JournalTitle{Frontiers in ICT}}}
  \textbf{\bibinfo{volume}{4}}, \bibinfo{pages}{29},
  \doiprefix\url{http://doi.org/10.3389/fict.2017.00029}
  (\bibinfo{year}{2017}).

\bibitem{hussain_optimal_2020}
\bibinfo{author}{Hussain, H.}, \bibinfo{author}{Javaid, M.~B.},
  \bibinfo{author}{Khan, F.~S.}, \bibinfo{author}{Dalal, A.} \&
  \bibinfo{author}{Khalique, A.}
\newblock \bibinfo{journal}{\bibinfo{title}{Optimal control of traffic signals
  using quantum annealing}}.
\newblock {\emph{\JournalTitle{Quantum Information Processing}}}
  \textbf{\bibinfo{volume}{19}}, \bibinfo{pages}{312},
  \doiprefix\url{http://doi.org/10.1007/s11128-020-02815-1}
  (\bibinfo{year}{2020}).

\bibitem{feld_hybrid_2019}
\bibinfo{author}{Feld, S.} \emph{et~al.}
\newblock \bibinfo{journal}{\bibinfo{title}{A {Hybrid} {Solution} {Method} for
  the {Capacitated} {Vehicle} {Routing} {Problem} {Using} a {Quantum}
  {Annealer}}}.
\newblock {\emph{\JournalTitle{Frontiers in ICT}}}
  \textbf{\bibinfo{volume}{6}}, \bibinfo{pages}{13},
  \doiprefix\url{http://doi.org/10.3389/fict.2019.00013}
  (\bibinfo{year}{2019}).

\bibitem{ding_implementation_2021}
\bibinfo{author}{Ding, Y.}, \bibinfo{author}{Chen, X.},
  \bibinfo{author}{Lamata, L.}, \bibinfo{author}{Solano, E.} \&
  \bibinfo{author}{Sanz, M.}
\newblock \bibinfo{journal}{\bibinfo{title}{Implementation of a {Hybrid}
  {Classical}-{Quantum} {Annealing} {Algorithm} for {Logistic} {Network}
  {Design}}}.
\newblock {\emph{\JournalTitle{SN Computer Science}}}
  \textbf{\bibinfo{volume}{2}}, \bibinfo{pages}{68},
  \doiprefix\url{http://doi.org/10.1007/s42979-021-00466-2}
  (\bibinfo{year}{2021}).

\bibitem{venturelli_quantum_2016}
\bibinfo{author}{Venturelli, D.}, \bibinfo{author}{Marchand, D. J.~J.} \&
  \bibinfo{author}{Rojo, G.}
\newblock \bibinfo{journal}{\bibinfo{title}{Quantum {{Annealing
  Implementation}} of {{Job-Shop Scheduling}}}}.
\newblock {\emph{\JournalTitle{arXiv:1506.08479 [quant-ph]}}}
  (\bibinfo{year}{2016}).
\newblock \eprint{https://arxiv.org/abs/1506.08479}.

\bibitem{nishimura_item_2019}
\bibinfo{author}{Nishimura, N.}, \bibinfo{author}{Tanahashi, K.},
  \bibinfo{author}{Suganuma, K.}, \bibinfo{author}{Miyama, M.~J.} \&
  \bibinfo{author}{Ohzeki, M.}
\newblock \bibinfo{journal}{\bibinfo{title}{Item {Listing} {Optimization} for
  {E}-{Commerce} {Websites} {Based} on {Diversity}}}.
\newblock {\emph{\JournalTitle{Frontiers in Computer Science}}}
  \textbf{\bibinfo{volume}{1}}, \bibinfo{pages}{2},
  \doiprefix\url{http://doi.org/10.3389/fcomp.2019.00002}
  (\bibinfo{year}{2019}).

\bibitem{IdeMaximumLikelihoodChannel2020}
\bibinfo{author}{Ide, N.}, \bibinfo{author}{Asayama, T.},
  \bibinfo{author}{Ueno, H.} \& \bibinfo{author}{Ohzeki, M.}
\newblock \bibinfo{title}{Maximum {{Likelihood Channel Decoding}} with
  {{Quantum Annealing Machine}}}.
\newblock In \emph{\bibinfo{booktitle}{2020 {{International Symposium}} on
  {{Information Theory}} and {{Its Applications}} ({{ISITA}})}},
  \bibinfo{pages}{91--95} (\bibinfo{year}{2020}).

\bibitem{Arai2021code}
\bibinfo{author}{Arai, S.}, \bibinfo{author}{Ohzeki, M.} \&
  \bibinfo{author}{Tanaka, K.}
\newblock \bibinfo{journal}{\bibinfo{title}{Mean field analysis of reverse
  annealing for code-division multiple-access multiuser detection}}.
\newblock {\emph{\JournalTitle{Phys. Rev. Research}}}
  \textbf{\bibinfo{volume}{3}}, \bibinfo{pages}{033006},
  \doiprefix\url{http://doi.org/10.1103/PhysRevResearch.3.033006}
  (\bibinfo{year}{2021}).

\bibitem{Yonaga2020}
\bibinfo{author}{Yonaga, K.}, \bibinfo{author}{Miyama, M.~J.} \&
  \bibinfo{author}{Ohzeki, M.}
\newblock \bibinfo{journal}{\bibinfo{title}{Solving {{Inequality-Constrained
  Binary Optimization Problems}} on {{Quantum Annealer}}}}.
\newblock {\emph{\JournalTitle{arXiv:2012.06119 [quant-ph]}}}
  (\bibinfo{year}{2020}).
\newblock \eprint{https://arxiv.org/abs/2012.06119}.

\bibitem{Koshikawa2021}
\bibinfo{author}{Koshikawa, A.~S.}, \bibinfo{author}{Ohzeki, M.},
  \bibinfo{author}{Kadowaki, T.} \& \bibinfo{author}{Tanaka, K.}
\newblock \bibinfo{journal}{\bibinfo{title}{Benchmark test of black-box
  optimization using d-wave quantum annealer}}.
\newblock {\emph{\JournalTitle{Journal of the Physical Society of Japan}}}
  \textbf{\bibinfo{volume}{90}}, \bibinfo{pages}{064001},
  \doiprefix\url{http://doi.org/10.7566/JPSJ.90.064001} (\bibinfo{year}{2021}).

\bibitem{Oshiyama2022}
\bibinfo{author}{Oshiyama, H.} \& \bibinfo{author}{Ohzeki, M.}
\newblock \bibinfo{journal}{\bibinfo{title}{Benchmark of quantum-inspired
  heuristic solvers for quadratic unconstrained binary optimization}}.
\newblock {\emph{\JournalTitle{Scientific Reports}}}
  \textbf{\bibinfo{volume}{12}}, \bibinfo{pages}{2146},
  \doiprefix\url{http://doi.org/10.1038/s41598-022-06070-5}
  (\bibinfo{year}{2022}).

\bibitem{Yamamoto2020}
\bibinfo{author}{Yamamoto, M.}, \bibinfo{author}{Ohzeki, M.} \&
  \bibinfo{author}{Tanaka, K.}
\newblock \bibinfo{journal}{\bibinfo{title}{Fair sampling by simulated
  annealing on quantum annealer}}.
\newblock {\emph{\JournalTitle{Journal of the Physical Society of Japan}}}
  \textbf{\bibinfo{volume}{89}}, \bibinfo{pages}{025002},
  \doiprefix\url{http://doi.org/10.7566/JPSJ.89.025002} (\bibinfo{year}{2020}).

\bibitem{Maruyama2021}
\bibinfo{author}{Maruyama, N.}, \bibinfo{author}{Ohzeki, M.} \&
  \bibinfo{author}{Tanaka, K.}
\newblock \bibinfo{journal}{\bibinfo{title}{Graph minor embedding of degenerate
  systems in quantum annealing}}.
\newblock {\emph{\JournalTitle{arXiv:2110.10930 [quant-ph]}}}
  (\bibinfo{year}{2021}).
\newblock \eprint{https://arxiv.org/abs/2110.10930}.

\bibitem{Amin2018}
\bibinfo{author}{Amin, M.~H.}, \bibinfo{author}{Andriyash, E.},
  \bibinfo{author}{Rolfe, J.}, \bibinfo{author}{Kulchytskyy, B.} \&
  \bibinfo{author}{Melko, R.}
\newblock \bibinfo{journal}{\bibinfo{title}{Quantum boltzmann machine}}.
\newblock {\emph{\JournalTitle{Phys. Rev. X}}} \textbf{\bibinfo{volume}{8}},
  \bibinfo{pages}{021050},
  \doiprefix\url{http://doi.org/10.1103/PhysRevX.8.021050}
  (\bibinfo{year}{2018}).

\bibitem{Kumar2018}
\bibinfo{author}{Kumar, V.}, \bibinfo{author}{Bass, G.},
  \bibinfo{author}{Tomlin, C.} \& \bibinfo{author}{Dulny, J.}
\newblock \bibinfo{journal}{\bibinfo{title}{Quantum annealing for combinatorial
  clustering}}.
\newblock {\emph{\JournalTitle{Quantum Information Processing}}}
  \textbf{\bibinfo{volume}{17}}, \bibinfo{pages}{39},
  \doiprefix\url{http://doi.org/10.1007/s11128-017-1809-2}
  (\bibinfo{year}{2018}).

\bibitem{Adachi2015}
\bibinfo{author}{Adachi, S.~H.} \& \bibinfo{author}{Henderson, M.~P.}
\newblock \bibinfo{journal}{\bibinfo{title}{Application of {{Quantum
  Annealing}} to {{Training}} of {{Deep Neural Networks}}}}.
\newblock {\emph{\JournalTitle{arXiv:1510.06356 [quant-ph, stat]}}}
  (\bibinfo{year}{2015}).
\newblock \eprint{https://arxiv.org/abs/1510.06356}.

\bibitem{benedetti2016estimation}
\bibinfo{author}{Benedetti, M.}, \bibinfo{author}{{Realpe-G{\'o}mez}, J.},
  \bibinfo{author}{Biswas, R.} \& \bibinfo{author}{{Perdomo-Ortiz}, A.}
\newblock \bibinfo{journal}{\bibinfo{title}{Estimation of effective
  temperatures in quantum annealers for sampling applications: {{A}} case study
  with possible applications in deep learning}}.
\newblock {\emph{\JournalTitle{Physical Review A}}}
  \textbf{\bibinfo{volume}{94}}, \bibinfo{pages}{022308},
  \doiprefix\url{http://doi.org/10.1103/PhysRevA.94.022308}
  (\bibinfo{year}{2016}).

\bibitem{Arai2021}
\bibinfo{author}{Arai, S.}, \bibinfo{author}{Ohzeki, M.} \&
  \bibinfo{author}{Tanaka, K.}
\newblock \bibinfo{journal}{\bibinfo{title}{Teacher-student learning for a
  binary perceptron with quantum fluctuations}}.
\newblock {\emph{\JournalTitle{Journal of the Physical Society of Japan}}}
  \textbf{\bibinfo{volume}{90}}, \bibinfo{pages}{074002},
  \doiprefix\url{http://doi.org/10.7566/JPSJ.90.074002} (\bibinfo{year}{2021}).

\bibitem{Sato2021}
\bibinfo{author}{Sato, T.}, \bibinfo{author}{Ohzeki, M.} \&
  \bibinfo{author}{Tanaka, K.}
\newblock \bibinfo{journal}{\bibinfo{title}{Assessment of image generation by
  quantum annealer}}.
\newblock {\emph{\JournalTitle{Scientific Reports}}}
  \textbf{\bibinfo{volume}{11}}, \bibinfo{pages}{13523},
  \doiprefix\url{http://doi.org/10.1038/s41598-021-92295-9}
  (\bibinfo{year}{2021}).

\bibitem{chancellor_modernizing_2017}
\bibinfo{author}{Chancellor, N.}
\newblock \bibinfo{journal}{\bibinfo{title}{Modernizing quantum annealing using
  local searches}}.
\newblock {\emph{\JournalTitle{New Journal of Physics}}}
  \textbf{\bibinfo{volume}{19}}, \bibinfo{pages}{023024},
  \doiprefix\url{http://doi.org/10.1088/1367-2630/aa59c4}
  (\bibinfo{year}{2017}).

\bibitem{ohkuwa_reverse_2018}
\bibinfo{author}{Ohkuwa, M.}, \bibinfo{author}{Nishimori, H.} \&
  \bibinfo{author}{Lidar, D.~A.}
\newblock \bibinfo{journal}{\bibinfo{title}{Reverse annealing for the fully
  connected \$p\$-spin model}}.
\newblock {\emph{\JournalTitle{Physical Review A}}}
  \textbf{\bibinfo{volume}{98}}, \bibinfo{pages}{022314},
  \doiprefix\url{http://doi.org/10.1103/PhysRevA.98.022314}
  (\bibinfo{year}{2018}).

\bibitem{noauthor_reverse_2017}
\bibinfo{title}{Reverse {Quantum} {Annealing} for {Local} {Refinement} of
  {Solutions}}.
\newblock \bibinfo{type}{Tech. Rep.}, \bibinfo{institution}{D-Wave Systems
  Inc.} (\bibinfo{year}{2017}).

\bibitem{Kadowaki2019}
\bibinfo{author}{Kadowaki, T.} \& \bibinfo{author}{Ohzeki, M.}
\newblock \bibinfo{journal}{\bibinfo{title}{Experimental and theoretical study
  of thermodynamic effects in a quantum annealer}}.
\newblock {\emph{\JournalTitle{Journal of the Physical Society of Japan}}}
  \textbf{\bibinfo{volume}{88}}, \bibinfo{pages}{061008},
  \doiprefix\url{http://doi.org/10.7566/JPSJ.88.061008} (\bibinfo{year}{2019}).

\bibitem{golden_reverse_2021}
\bibinfo{author}{Golden, J.} \& \bibinfo{author}{O’Malley, D.}
\newblock \bibinfo{journal}{\bibinfo{title}{Reverse annealing for
  nonnegative/binary matrix factorization}}.
\newblock {\emph{\JournalTitle{PLOS ONE}}} \textbf{\bibinfo{volume}{16}},
  \bibinfo{pages}{e0244026},
  \doiprefix\url{http://doi.org/10.1371/journal.pone.0244026}
  (\bibinfo{year}{2021}).

\bibitem{noauthor_gurobi_2020}
\bibinfo{organization}{Gurobi Optimization, LLC.}
\newblock \emph{\bibinfo{title}{Gurobi {Optimizer} {Reference} {Manual}}}
  (\bibinfo{year}{2020}).

\bibitem{land_automatic_1960}
\bibinfo{author}{Land, A.~H.} \& \bibinfo{author}{Doig, A.~G.}
\newblock \bibinfo{journal}{\bibinfo{title}{An {Automatic} {Method} of
  {Solving} {Discrete} {Programming} {Problems}}}.
\newblock {\emph{\JournalTitle{Econometrica}}} \textbf{\bibinfo{volume}{28}},
  \bibinfo{pages}{497--520}, \doiprefix\url{http://doi.org/10.2307/1910129}
  (\bibinfo{year}{1960}).

\bibitem{mcgeochDWaveAdvantageSystem}
\bibinfo{author}{McGeoch, C.} \& \bibinfo{author}{Farre, P.}
\newblock \bibinfo{title}{The {{D-Wave Advantage System}}: {{An Overview}}}.
\newblock \bibinfo{type}{Tech. Rep.}, \bibinfo{institution}{D-Wave Systems
  Inc.} (\bibinfo{year}{2020}).

\bibitem{Ohzeki2019JPSJ}
\bibinfo{author}{Ohzeki, M.}
\newblock \bibinfo{journal}{\bibinfo{title}{Message-passing algorithm of
  quantum annealing with nonstoquastic hamiltonian}}.
\newblock {\emph{\JournalTitle{Journal of the Physical Society of Japan}}}
  \textbf{\bibinfo{volume}{88}}, \bibinfo{pages}{061005},
  \doiprefix\url{http://doi.org/10.7566/JPSJ.88.061005} (\bibinfo{year}{2019}).

\end{thebibliography}

\section*{Acknowledgments}
\addcontentsline{toc}{section}{Acknowledgments}
The authors would like to express sincere gratitude to Assistant Prof. Manaka Okuyama for fruitful discussions and kind support for this study.
This work was financially supported by JSPS KAKENHI Grant Nos. 20H02168, 19H01095, and 18H03303, partly supported by JST-CREST (No. JPMJCR1402), the Next Generation High-Performance Computing Infrastructures and Applications R\&D Program of MEXT, and by MEXT-Quantum Leap Flagship Program Grant Number JPMXS0120352009.

\section*{Author contributions statement}
\addcontentsline{toc}{section}{Author contributions statement}
R.H. conceived of the presented idea and performed the experiments.
M.O. and K.T. verified the analytical methods and supervised the findings of this work. All authors discussed the results and contributed to the final manuscript.

\end{document}